# Predicting the Stock Price of Frontier Markets Using Modified Black-Scholes Option Pricing Model and Machine Learning

Reaz Chowdhury[1], M.R.C. Mahdy[1,2*], Tanisha Nourin Alam[1], Golam Dastegir Al Quaderi[3]

[1]*Department of Electrical & Computer Engineering, North South University, Bashundhara, Dhaka 1229, Bangladesh*

[2]*Pi Labs Bangladesh LTD, ARA Bhaban,39, Kazi Nazrul Islam Avenue, Kawran Bazar, Dhaka 1215, Bangladesh*

[3] *Department of Physics, University of Dhaka, Dhaka 1000, Bangladesh*

*Corresponding Author: mahdy.chowdhury@northsouth.edu

**Abstract**

The Black-Scholes Option pricing model (BSOPM) has long been in use for valuation of equity options to find the prices of stocks. In this work, using BSOPM, we have come up with a comparative analytical approach and numerical technique to find the price of call option and put option and considered these two prices as buying price and selling price of stocks of frontier markets so that we can predict the stock price (close price). Changes have been made to the model to find the parameters 'strike price' and the 'time of expiration' for calculating stock price of frontier markets. To verify the result obtained using modified BSOPM we have used machine learning approach using the software Rapidminer, where we have adopted different algorithms like the decision tree, ensemble learning method and neural network. It has been observed that, the prediction of close price using machine learning is very similar to the one obtained using BSOPM. Machine learning approach stands out to be a better predictor over BSOPM, because Black-Scholes-Merton equation includes risk and dividend parameter, which changes continuously. We have also numerically calculated volatility. As the prices of the stocks goes high due to overpricing, volatility increases at a tremendous rate and when volatility becomes very high market tends to fall, which can be observed and determined using our modified BSOPM. The proposed modified BSOPM has also been explained based on the analogy of Schrodinger equation (and heat equation) of quantum physics.

## Introduction

The quest for a valuation formula that would represent option pricing had first started at the beginning of twentieth century by Louis Bachelier. In 1960s, significant progress had been made on valuation of option pricing, which were reflected on the great works done by some financial economists. One of the most basic and powerful tools in financial mathematics is the Black-Scholes equation, which is used to find the prices of assets. The price of European call options can be determined as the solutions of Black-Scholes equation [1, 2]. Moreover, to model transaction costs arising in the hedging of portfolios [3, 4] and large traders' feedback effects [5-8], non- linear Black-Scholes equation have been derived. An option is defined to be a contract or security that gives the buyer or holder the right but not the obligation to buy or sell assets at a specified price called strike price and at a specified date called expiration date of the stock [9, 10]. In option pricing, if the price of the stock is higher than the strike price then the option tends to have higher value and so the option is exercised but if the price of the stock is less than the strike price then the option has less value and so the option is given to expire [9].

Historically, options are one of the most complicated and abstruse fields among financial instruments because of its uncertainty and it is a very perplexing problem in finance. However, with the developments of option pricing model, it has now became easier to trade options and so option pricing is no longer viewed as esoteric financial instrument . Option can be of two types- call and put. Call option is said to be in the money if the strike price is below the current stock price, at the money if the current stock price is near or at the strike price and out of the money if the stock price is below the strike price. In case of put options, it is said be in the money if the current stock price is below the strike price, at the money if the strike price is at or near the current stock price and out of the money if the stock price is above the strike price. However, in reality this is not always what happens in case of exercising the options and the reality is much more deviated from the theoretical explanation of the BSOPM. The deviation occurs because there is risk in trading and factors like supply and demand, bank rate, speculative pressure etc., whereas Black-Scholes developed a model in which they argued to value options by removing "risk" parameter through dynamic hedging [11]. So including the risk back in means deviation of the theoretical price from the real trading price.

Stock price prediction of a particular company of frontier markets using Black-Scholes formula have not been studied so far. It is because frontier markets do not follow the method of an emerging market, where trading of options takes place and that the asset price like stock price can be predicted using BSOPM. The reason is that BSOPM has parameters called strike price and time to maturity and volatility, which do not even exist when it comes to trading in frontier markets. However, machine learning and data mining play a very important role, when it comes to predicting stock prices of both frontier and derivative markets. Applications of machine leaning and data mining techniques have led to the growth of various business intelligence systems in recent decades. It is because algorithms can be trained and developed by using machine learning so that it can almost accurately predict the future stock prices based on historical prices of the stocks and can also display the results through prediction and description. On the other hand, as the Black-Scholes option pricing model had been drawn quite heavily from physics based notion of Brownian motion so a number of different approaches have been constructed and established to derive the Black-Scholes partial differential equation in the light of quantum mechanics.

In order to predict the stock prices of frontier markets using Black-Scholes option pricing model, we have developed an approach by simply changing and modifying the parameters "strike price", "time of maturity" and "volatility" of the Black-Scholes equation. As in frontier markets trading goes on continuously so we have followed the "American option" trading method, where options can be exercised at any period of time and so it is more compatible for predicting the stock price of frontier markets. One of the interesting findings of this research is related to the parameter "volatility" of Black-Scholes equation. It has been found that, the market crash is related directly with the sudden huge increase in volatility of the stocks. If the volatility increases continuously, month by month along with the increase in the stock price of companies of a frontier market like Dhaka stock exchange at a tremendous rate, then it has to be understood that, the market is likely to get crashed or fall in near future. It is because the volatility can not remain high for a long period of time due to overpricing as people do not trade stocks in the same way everyday and so there have to be some increment and decrement in the prices of the stocks in the market. To observe the increase and decrease of volatility, we have used MATLAB. It is because the software takes into account the close price (stock price) of the stock of a particular company of a frontier market for calculating the volatility and displays the volatility based on this close price of the stock with the corresponding date on which the stock was traded and hence, it becomes easier for a trader to see the change in volatility of the stock. This can be very useful for a trader to observe the trend of the market and ending up making some profit out of the stocks. The results are obtained using Black-Scholes equations in such a way that it can illustrate the comparison between original stock price and predicted close price of particular stocks of frontier markets. In order to verify the results obtained using Black-Scholes equations, we have used data mining and machine learning techniques in Rapidminer, where we have used some algorithms and methods like decision tree, neural net, linear regression and ensemble learning method to predict the future stock prices of eleven companies of different stock exchanges of frontier markets. The changes and modifications made to the Black-Scholes equations have also been explained from the point of view of quantum physics based on the analogy of Schrodinger equation.

**Black-Scholes-Merton Equation**

Black and Scholes derived an option pricing formula using which the theoretical value of options can be determined [9]. However, the model was based on some certain parameters having some assumptions and later, R.C Merton further extended the model for dividend payments and exercise price changes [1]. The formulas are given below-

$$c = Se^{-qT}N(d_1) - Xe^{-rT}N(d_2) \tag{1}$$
$$p = Xe^{-rT}N(-d_2) - Se^{-qT}N(-d_1) \tag{2}$$

where,

$$d_1 = \frac{\ln\left(\frac{S}{X}\right) + \left(r - q + \frac{\sigma^2}{2}\right)T}{\sigma\sqrt{T}} \tag{3}$$

$$d_2 = \frac{\ln\left(\frac{S}{X}\right) + \left(r - q - \frac{\sigma^2}{2}\right)T}{\sigma\sqrt{T}} \tag{4}$$

Here,

$S = Stock\ Price$

$X = Strike\ Price$

$r = Risk\ free\ interest\ rate$

$T = Time\ to\ expiration\ in\ years$

$\sigma = Volatility\ of\ the\ relative\ price\ change\ of\ the\ underlying\ stock\ price$

$N(x) = Cumulative\ normal\ distribution\ function$

However, this equation only deals correctly with European options because European options hold an expiration date unlike American options. In this paper, we will follow the American options pricing method to find the stock prices of the frontier markets using a different approach of Black-Scholes-Merton model by changing some parameters to better work in the frontier markets.

**Methodology**

BSOPM assumes that the stock price distribution is lognormal at any finite interval of time as the stock price in BSOPM follows a random walk in continuous time, where price of the stock is proportional to the variance rate [9]. BSOPM assumes the price of assets follows a geometric Brownian motion with constant drift and volatility. The first solution to find a fair and riskless price for instruments like options was given by Bachelier in 1900 [12], who first used Brownian motion for mathematical modeling of stock price movements, where he used central limit theorem to derive a normal distribution of stock price movements [13]. So for each day's stock price, we can write-

$$Today'Stock\ Price = Yesterday's\ Stock\ Price * e^u$$

Where, 'e' is an exponential term function and 'u' is defined to be the periodic daily rate of return. It is the rate, at which the asset increased or decreased that day. Because the periodic rate of return on an asset is a random number, so to shape the movement and determine the future stock price we have used a formula that models a random movement. This is mainly a stochastic process that Bachelier used to describe changes in the prices of the stocks, which is now known to us as Brownian motion [13]. Brownian motion assumes there are two parts of an asset, whereas BSOPM considers that the variance rate of the return on the stock is kept constant having the interest rate constant and known. So together for defining the periodic daily rate of return [u] based on BSOPM and the Brownian motion we can write-

$$u = \left(r - \frac{\sigma^2}{2}\right)t + \sigma w_t \tag{5}$$

Here, the first part is a constant drift and the second part is random stochastic component. So the future stock price is,

$$Future\ Stock\ Price = Current\ Stock\ Price * e^{\left(r - \frac{\sigma^2}{2}\right)t + \sigma w_t}$$

The stock will thus go up in a constant rate over time or at the risk free rate if the risk is removed from the stock. But as people in frontier markets randomly buy and sell stocks so it has higher volatility and so including the risk back in means including the implied volatility. That means, each day the asset can increase or decrease randomly. So according to the central limit theorem if these large number of simple random samples of a particular company is taken from the population and the mean is calculated from each then the distribution of these sample means will assume the normal probability distribution. In other words, if we make a graph of periodic daily return then the graph will form a normal distribution bell shaped graph. So we assume that the rates of daily change of price in the future will also be normally distributed. Thus, the graph of future periodic rates of return is a normal distribution curve where the drift is the mean and the historical future standard deviation as the assumed future standard deviation, which is what Brownian motion in Black-Scholes formula is. The Brownian motion means if we graph the future periodic rate of return we assume that, the graph will form a normal distribution bell shaped curve, using drift as the mean and using historical standard deviation as the future standard deviation.

$$Mean, [\mu] = r - \frac{\sigma^2}{2}$$

The total probability of what the future rate of return is represented by normal distribution curve and so we can easily use this to find the probability that the stock price will be above or below the strike price of an option on the expiry date. So we find the rate of growth that it would take for the current stock price to be at the strike price when the option expires and see where the rate of growth falls onto the curve of normal distribution. To do this we find how many standard deviation away the rate of growth from the stock price to the strike price is from the expected rate of growth. This is known as the standard Z score. The Z score is probability and statistics is defines as-

$$z = \frac{X - \mu}{\sigma}$$

Where,

$X = ln\left(\frac{K}{S}\right) = The\ rate\ of\ growth\ from\ stock\ price\ to\ the\ strike\ price$

$\mu = \left(r - \frac{\sigma^2}{2}\right) = Mean\ of\ probability\ curve\ that\ represents\ the\ future\ rate\ of\ growth$

$\sigma = The\ standard\ deviation\ of\ the\ periodic\ daiy\ return\ over\ an\ one\ year\ period$

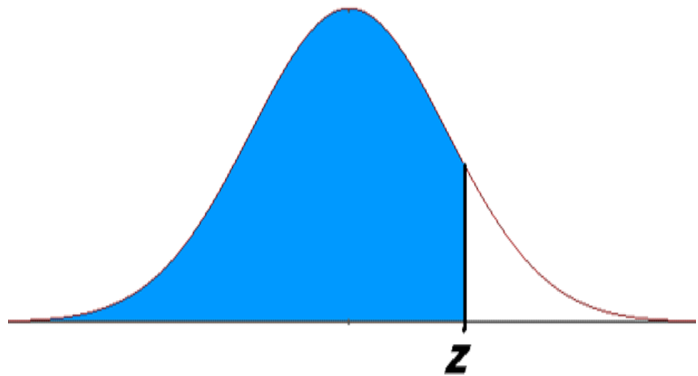

Figure 1: Figure of a probability distribution curve.

In figure 1, the total area under the curve represents the total probability of what the future rate of growth will be. The percent area that lies to the right of Z score represents the probability that the stock price will be at or above the strike price, when the option expires, whereas the left area represents the probability the stock price being below the strike price when option expires.

**Finding "strike price" parameter**

The fixed price of an option at which the owner of an option can buy or sell the underlying security is the strike price [9]. Strike price is one of the most important factors when it comes to value an option because without the strike price it is not possible to determine whether the option is valuable or worthless. The problem in frontier markets is that there is no strike price as the market doesn't trade options but stock shares. So to find the strike price to better work in the frontier markets, we have calculated it mathematically, which is derived below depending on the lognormal distribution described above in methodology.

If $X = \ln\left(\frac{K}{S}\right)$ then we have,

$$e^X = \left(\frac{K}{S}\right) \qquad (6)$$

and so we can write,

$$Strike\ price\ input\ in\ BS\ equation, K = (S * e^X) \qquad (7)$$

We can find the daily price change (X), which is for a current day the natural log of today's close price of the stock divided by yesterday's close price of the stock for a particular stock of a company. So,

$$X = \ln\left(\frac{Today's\ close\ price}{Yesterday's\ close\ price}\right)$$

Then averaging all daily price changes of a particular month [see table 7] we get the mean which is kept constant by changing all other variables and thus call and put options are calculated [see table 8] in order to find the buying and selling price of the market. So this is how we can find the strike price for the current day, which is observed to be in the money, at the money and also out of the money similar to the option pricing method.

**Finding "Time" parameter**

Time of expiration is one of the most important key factors, which has a significant effect in the stock price in BSOPM. With the passing of time, option price also decreases if the price of the stock does not change [9]. It is important to note that, the time is in years (theoretically) but option traders also work with days remaining expiration.

The problem in frontier markets is that there is no time of expiration, as it has no options to trade. So as stock shares do not have an expiry time, people trade stocks randomly and continuously, which actually is a never ending process. People mainly trade stock shares of companies based on certain parameters like last trade price (LTP), high, low, opening price, close price, value, dividend, volume, EPS (earnings per share), P/E ratio (price earnings ratio), long and short time loan of the company, market capital etc. So a buyer can buy stocks of a certain company and can hold it throughout his or her lifetime and can trade at any time at any price. To adjust the time of expiration, we have taken into account the trading days over the year. So it is calculated as the exact number or the trading day of a month over the total number trading days of the whole year. In this paper, we have considered this approach to predict the price of a stock in future at any day. For example- if a buyer wants to buy a stock of company Afric Industries SA in 1st March 2016 then the exact number of calendar day of 1st March, 2016 is taken into consideration. Let us consider the following condition- starting from the month January 2016 up to the month of February 2016 trading took place for 42 days in Casablanca Stock Exchange so the 43rd day of course is the 1st March, 2016. The total number of trading days up to the last day of trading in month of February 2016 starting from the back in the first day of month March 2015 will be a year, which approximately is 261 days in total. So the time for stock price in 1st March will be-

$$Time\ input\ in\ BS\ equation = \frac{43}{261} = 0.164 \qquad (8)$$

So the time is taken to be simply to be the exact number of trading date on which the share will be traded over the total number of trading days of the year. However, the denominator side is kept constant but the numerator will change. So in this way by changing the time continuously along with stock price of any day, the value of call and put options are determined.

**Finding 'Volatility' parameter**

The parameter that plays the central role in derivative market is the volatility as it is directly related to the movement of the stock prices. With the increase of the volatility, range of share prices becomes much wider than the range of a low volatility stock. Options are non-linear securities, which are traded by the traders having the volatility information [14]. Thus, traders make profits by taking advantage of stocks, which moves at higher margins due to higher volatility but this also includes high risk in the asset's price. This volatility is of two types- one is historical volatility and another one is implied volatility. Implied volatility of the Black-Scholes model is the value of the volatility that makes the Black-Scholes price match the market price. That means, the implied volatility comes from the market. The historical volatility is the one, which is derived from time series of past market prices. But it is not possible to know the future volatility of the stocks for trading and so traders use historical volatility to trade options.

The problem in frontier market is there is no volatility mentioned in different stock exchanges like Dhaka Stock Exchange (DSE), Colombo Stock Exchange (CSE), Pakistan Stock Exchange (PSX), Casablanca Stock Exchange etc. So the volatility needs to be calculated in order to put it as an input to the Black-Scholes equation. Volatility is the standard deviation and can be determined by two methods- one is using the direct formula, which is-

$$\sigma = \sqrt{\frac{\sum_{i=1}^{N}(X_i - \mu)^2}{n - 1}}$$

Where,

$\sigma = standard\ deviation$
$X = daily\ price\ change\ of\ the\ stocks$
$\mu = mean\ or\ average\ of\ the\ stocks$
$n = number\ of\ values\ in\ the\ dataset$

The second method is using MATLAB Simulink. In this paper, we have used the MATLAB Simulink approach as it gives the standard deviation directly in its data statistics. To determine the volatility, the close price of a month is taken in the Y-axis and the corresponding dates are taken in the X-axis. Thus, MATLAB shows the min, max, mean, media, mode, standard deviation and range of the prices in data statistics located under tools. This is done with the co-relation of input parameter "time". As time is over one year of period, so the volatility is calculated for every month separately for each year. Finally, the volatilities of twelve months of one year are added together and thus provided as the input to the Black-Scholes equation [the MATLAB code to find the volatility of a stock over one-year period is given in supplement S1]. The chart and corresponding values of volatility of company Afric Industries SA (AFI) for one year obtained using MATLAB to predict call option and put option is shown in table 9.

**Long and short positions for frontier markets**

Long position is the one, which is associated with investing in the market as it is the expectation of an investor that the assets price will increase in future. Investors on the long position mainly banks on the increase of a price of an asset.

Short investing is quite the opposite of long position strategy because it involves borrowing and then selling a security with the expectation of the trader that the asset price will somehow decrease. Investors on short position banks on decrease of price of an asset. But there is risk in short selling because the buyer has to borrow a security from a broker and have to make profit out of it by selling the security in the open market. Then the buyer has to buy back the security after its price decreases below the price at which the buyer sold it in the market and then again sell the security back to the broker. Investors short a stock whenever they expect that the price will decline and so can make profit out of it by taking advantage of declining stocks. So as risk is involved in short selling, the investors prefer long position over short position because in long position the risk is much lower than short position as the share price can never fall below zero dollar.

Although in frontier markets there are no terms like long and short selling but in reality it happens. It's because traders in frontier markets actually buy stock shares and can hold it as long as the buyer wants. So the buyer can actually go for long position and short position depending on the trend of the market. But the fact is that the buyer never borrows the share from the broker in frontier market and so the buyer is not obligated to return the stock share after making profit. Using Black-Scholes-Merton equation, we have come up with an idea to determine the buying price of the stock and also the selling price of the stock. The "call option" and "put option" parameters which are actually the right to buy a security and right to sell a security hence becomes the buying price of a stock share and selling price of a stock share for a particular company in case of frontier markets. Then averaging the both market prices, the close price of the stock can be determined. This is because the close price of a stock of any company is usually calculated by averaging the last twenty market prices that takes place in the last few minutes before market closes [comparison of original close price and predicted close price of all eleven companies using Black-Scholes equation is given in supplement S3].

**Dividend Yield**

In this work, it has been observed from Black-Scholes-Merton equation that the price of any stock decreases if dividend yield increases. When a historical close price of a stock of any company is given as input in the Black-Scholes equation along with other parameters including dividend yield to find the theoretical close price of the stock of another day then it is seen that due to high dividend yield the theoretical close price decreases from the actual close price. That means, dividend paying stocks tends to have decreasing value of call and put options than the non-dividend paying stocks. As we have taken the average of call option and put option to be the close price of a particular stock, so when the value of call and put options decreases, the theoretical value deviates much from the original call and put values as a result of which the close price we obtain also deviates from the original close price. The dividend yield of all the stocks of the companies, which are given as input in the Black-Scholes equations are collected using the Thomson Reuters software and displayed in table 10. The dividend yield of all the companies' stocks are collected using Thomson Reuters, expresses the dividend per share as a percentage of the share price.

**Risk Free Interest Rate**

In reality there is no term called risk free interest rate but the Black-Scholes-Merton PDE assumes that the short term interest rate is known and constant through time [9]. That is why, we have collected all data using Thomson Reuters software considering the treasury bills to be the risk free interest rates for all companies given by their corresponding central banks of the respective countries, which is given in table 11.

**Machine learning approach**

Machine learning is one of the most important and fundamental approaches to play a wide range of critical roles when it comes to data mining, text mining, predictive analytics and decision making systems. Machine learning and data mining technique can be integrated into business intelligence systems to help making real life decision [15, 16-20]. Machine learning can be divided into two subcategories namely representation and generalization. We have approached towards generalization category as it predicts the accuracy of an unseen data based completely on training dataset [21].

In this paper, we have used the widely used software called Rapidminer to preprocess and analyze the stock prices as it supports all steps of data mining process [22]. Effort to predict the trend and pattern of stock market has been a very challenging one [23-27]. In Rapidminer software, data analysis is usually performed using graphs, plots, charts and tables in which one can easily visualize the output and also compare between one or more attributes and models. But for a machine to predict the future stock price, it is essential to train the machine to learn from the given dataset based on which models will be created from different algorithms and thus the prediction will be accomplished [21]. To do that, the models have to be developed by providing it with different algorithms and with a training dataset to learn and thus the output will be given based on probability distribution function and frequency of the dataset [21]. In Rapidminer platform for getting unexpected relationships from large dataset, data mining stands as one of the most powerful field of study [28].

Predicting stock price is one the most complicated areas of applied finance, where machine learning can play a vital role to predict the future stock prices. In order to maximize capital gain and minimizing loss to get the optimum output, there is a need for accurately predicting the trends in stock market prices [29]. We have used three learning algorithms in Rapidminer, which are used for creating prediction models in order to predict the future stock price of eleven companies [see table 1] and a large input set of data has been provided to train those models [reference of data taken from various websites are shown in supplement S4]. The datasets have been collected from four Stock Exchanges [see table 10]. Our prediction involves six attributes [see table 1] to analyze and predict the future stock price, where the close price is given as 'label' attribute and date as an 'ID'.

**Predictive Models**

**Decision Tree**

Decision tree algorithm is one of the most important tree models when it comes to prediction of stock prices. This operator generates a decision tree model that can be used for classification and regression [30]. This operator separates values belonging to different classes for classification and in case of regression it separates them in order to reduce the error in an optimal way for the selected parameter criterion. It can easily be learnt by dividing the main dataset into two subsets [22].

Rapidminer provides to generate decision tree model and it automatically generates the tree model to ensure better prediction for the close price of the companies we have experimented on. To do that, we have divided our dataset separately for testing and training in excel spreadsheet format [see table 2]. For predicting stock price of any company for a particular month based on previous data, we have provided historical prices as input to train the model along with the close price as label. For testing, we have provided data only for the month for which the prediction will be done. The model outlook is shown in Figure 3. This model learner works by forming tree like collection of nodes intend to create an estimation of numerical target value. In our case, this target value is close price of the stock.

The parameters of decision tree have been changed to increase and obtain the best accuracy for prediction of close price. The criterion parameter is set to least square on which the close price attribute is selected for splitting and minimizing the squared distance between the average of values in the node with regards to the true value. The maximum depth of the trees is tuned to 25 from default value of 20 because of the large input of dataset. The minimal gain is tuned to 0.15. The performance operator is used for statistical performance evaluation and it delivers a list of performance criteria values of the regression task. The performance vector obtained by using decision tree model is shown in table 3 for all eleven companies. The graphs obtained from rapidminer, which shows the comparison of original close price and predicted close price for all eleven companies is shown in figure 2 (a)-(k).

**Neural net**

An artificial neural network is the one defined as a system, which is modelled on the human brain and nervous system because it is an interconnected network, which works through various nodes called neurons. A large number of connected processing units works together to function and process the output information. Fewer the number of neurons, higher the performance of the system [22].

Rapidminer provides to generate neural network model. For performing neural network in Rapidminer, we have used the same model used in [21], which is given in figure 4. The dataset consists of given subsets, which are training and testing. The two datasets are the same as decision tree shown in table 2. For predicting stock price of the eleven companies through neural net, we have trained the model by giving historical prices of all six attributes as input. In the testing set, we have given the attributes of the particular month for which the prediction will be done. The set role operator works to set the role of the attribute date as an ID. Windowing operator is used to transform the given example set containing series data into a new example set containing single valued example. To work accordingly with the given input dataset, the parameters series representation, window size, step size are changed and set in a proper way. Series representation is defined as how the series values will be encoded and is set as encode series as example by default. Window size is the width of the used windows and step size is the distance between the first values. Both these parameters are set to have a value 1. Create single attributes and create label has been check boxed because we have focused on the close attribute that is, we are predicting the close attribute of the future depending on the close attribute of the past. The close attribute is set as label. So in training window operator a label attribute is created, which is the close price, whereas in testing side no label attribute is created as it will be predicted by the model. A validation (sliding window validation) operator is used for series prediction as the time points are encoded as examples. It uses both windows of examples, one window for training and uses another window for testing as shown in figure 4. The parameters for this operator are training window width, training window step size, test window width and horizon. Training window width is the number of examples in the window, which is used for training and tuned to 3. Training window step size is the number of examples the window is moved after each iteration and this parameter is tuned to 1. Test width window is the number of examples, which are used for testing and has been set to 3. Horizon is the increment from last training to first testing example and this value is tuned to 1. There are two phases in the sliding window validation operator, which are training phase and testing phase. In the training phase, the learning algorithm is neural net so that it can learn from the given training dataset. The number of training cycles used for neural net training is set to be 500 cycles. The learning rate is amount of change of weights at each step and is tuned to 0.3 whereas the momentum is tuned to 0.2, which simply adds a fraction of previous weight update to the current one. The testing phase contains the apply model operator and performance parameter (forecasting performance). A comparative chart is shown in figure 5 (a)-(k) to visualize the difference and deviation between the original and predicted close price and performance vector obtained using Rapidminer is shown in table 4. A sample of neural network is shown in supplement S2.

**Ensemble Method**

Ensemble learning is a machine learning technique, where multiple machine learning algorithms or learners strategically generated and combined together in order to solve one particular computational intelligence problem. In ordinary machine learning approaches, we try to draw one hypothesis from the given training data but in ensemble method, we can construct set of hypotheses and combine them. Ensemble learning improves the performance of a model or it reduces the likelihood of an unfortunate selection of a bad model for prediction. So ensemble learning method ensures the best prediction analysis and optimum output. An ensemble method contains a number of learners or classifiers, which are called base learners and the generalization ability of the ensemble is much stronger than the base learners. Base learners are usually generated from the training data by a base learning algorithm which can be decision tree, neural networks, relative regression algorithm or other kinds of machine learning algorithm. Most ensemble methods use only a single base learner to produce homogenous learners but there are some methods, which uses multiple base learners to make a heterogeneous learner that can make very accurate predictions [31]. Ensemble learning decreases the risk of making a particularly bad selection of algorithm or classifier and this method can really be useful when the problem deals with large dataset. There are many ensemble methods, which have been effective enough to produce accurate predictions like boosting, bagging and stacking.

In this paper, we have used ensemble method to get the optimum output from the given learning algorithm. In machine learning and data mining, it is believed that, a group of models is better than a single model, which is the main principle of ensemble learning. It helps to overcome the biases and error rate of the individual models, which is done by producing a strong learner combining some weak learners. As in case of ensemble learning, the training dataset plays the most effective contributor to the error in the model so we have taken a large training set along with six predefined attributes and close price as the label attribute. The model is shown in figure 6. The split data operator produces the desired number of subsets of the given example set. The example set is partitioned into subsets according to the specified relative sizes. So for splitting the given dataset, the partition parameter of this operator is tuned to have two partitions in a ratio of 0.6 (for training) and 0.4 (for testing). The sampling type is set as linear sampling.

The vote operator is an ensemble operator, which is also a nested operator that means it has a sub-process. This sub-process has three learners, which are decision tree, neural net and relative regression. The training dataset to build these classifiers is shown in table 5.

The parameters in the models are changed. In case of decision tree, the parameters are criterion, maximum depth. The role of criterion is to select the criterion on which attributes will be selected for splitting. We set this value as least square in which attribute is selected for splitting, that minimizes the squared distance between the average of values in the node with regards to the true value. The maximal depth is used to restrict the depth of the decision tree and it is tuned to 20. A sample leaf of tree diagram is shown in figure 7. For neural net, there are 500 training cycles, learning rate tuned to 0.3 and momentum tuned to 0.2. The relative regression model is a regression model, which is very useful in order to allow time series predictions on datasets with large trends because it learns a regression model for predictions relative to another attribute value. In our case, the attribute value is set as the close price. The relative regression contains a linear regression operator, which calculates a linear regression model from the input dataset, which is provided for training. However, it has been found that by using multiple relative regression model in the vote sub process the accuracy of the overall performance of the models is increased. The overall prediction of this ensemble will be the majority of the prediction by the individual base classifiers. So to get the majority votes of the individual base learners, apply model operator has been set in building the model along with the performance operator, which will determine the performance (accuracy). The performance vector is shown in table 6. Finally, comparison charts [cf. figure 8 (a)-(k)] are shown to visualize the difference and deviation between original and predicted close price of the stock, whereas in figure 9 (a)-(k), the comparison charts of all machine learning and Black-Scholes predictions with the original close prices are given.

### Explanation of the Modification of BSOPM from the Point of View of Quantum Physics

Different approaches could be used to derive the Black and Scholes PDE, which also include those employing the concepts of quantum physics [32-37] and other branches of physics. Physicists are continuously trying to better understand the financial dynamics using physical ideas and concepts [38-42]. In this part, we would mainly like to focus on the logical reasons behind the changes made to the parameter named 'time of expiry' and 'strike price' and why these have been changed so.

The option price depends on the stock price and this stock price is a random variable evolving with time [43, 32]. The option pricing method introduced by Black, Scholes and Merton is composed of stochastic differential equation, which is however within the interpretation of Ito calculus. The Black and Scholes PDE can be written as a diffusion equation in the following form:

$$\delta_t \pi = -\frac{1}{2}(\sigma s)^2 \delta_{ss}\pi - rs\delta_s\pi + r\pi \tag{9}$$

The root idea behind the formulation of Black-Scholes equation is standard Brownian motion process, which is given by-

$$ds(t) = \mu s(t)dt + \sigma s(t)dW(t) \tag{10}$$

In (10) W is a wiener process or standard Brownian motion process, which is actually a continuous stochastic process. In this paper we also have a random stochastic component in equation (5), which is used to find the strike price of the corresponding stock price.

Furthermore, the random term (9) in the stochastic equation must be delta-correlated as the market incorporates instantaneously any information concerning future market evolution, which is efficient market hypothesis [32, 43- 45]. This implies that, theoretical prices are driven by white noise [32], which is a combination of white shot noise and wiener process [46]. BSOPM have been extended to introduce white shot noise [47, 48]. However, in discrete time, the samples of white noise are sequences of uncorrelated random variable with zero mean and finite variance and the samples are independent having identical probability distributions. But we know that if each sample has normal distribution having zero mean then the signal is a Gaussian white noise. However, in this research, we have found that the mean is not always zero [see table 7] and is slightly more than zero, which concludes that there is a finite chance that white noise may not exists in BSOPM. Moreover, as the samples are sequential in time, the stock prices in Black-Scholes-Merton equation are also sequential in time. Hence, the parameter 'time' is changed sequentially in this research, that is, time of expiration also sequentially changes along with the stock price. Hence, the time of expiration is dependent on the particular day's stock price and varies accordingly with the stock price of a certain day. The implied volatility also varies over time and at any point in time over related options, the characteristic of the options differ only by the strike price [49].

Thirdly, the Black-Scholes equation obtained using Stratonovich calculus and Ito calculus are the same, which means that time does not change when different approaches are considered [43]. It changes only with the stock price of a certain day. In this paper, we have considered the American option pricing method in which an option can be exercised at any time unlike the European option pricing method. However, the future price of a commodity or asset is the price at which one can agree to buy or sell it at a given time in the future [50]. Since, the future price is a function of time, upon which the asset would be bought or sold so, the time is dependent on that particular day's stock price. So, time is taken simply to be the time or date of a particular day on which the stock will be traded.

The function $\pi = \pi(s, t)$ represents the option price as a function of time and asset price. Changing coordinates the Black-Scholes model is portrayed [33] in the following way-

$$\frac{\partial \pi}{\partial t} + \frac{1}{2}\sigma^2 \frac{\partial^2 \pi}{\partial x^2} = 0 \tag{11}$$

Making a Wick rotation in time $t = -i\tau$, the heat equation (11) becomes

$$i\frac{\partial \pi}{\partial \tau} = -\frac{1}{2}\sigma^2 \frac{\partial^2 \pi}{\partial x^2} \tag{12}$$

Equation (12) is the Schrodinger equation [33] of a free particle with $\hbar = 1$ and $m = \frac{1}{\sigma^2}$. If we compare this equation with Schrodinger equation, then it is clearly visible that the left side of equation (12) contains imaginary number $i$ and the Wick-rotated 'time' $\tau$, which are also imaginary. That means, the function $\pi = \pi(\tau, x)$, changing with imaginary 'time' $\tau$ represents an option, which will be traded in a definite given time in future. This is a valid reason for changing time along with the change of the stock price. Again, if we compare this equation to the heat equation, the function $\pi = \pi(\tau, x)$ that describes the option at a given location $x$ will change over time, as heat will spread throughout space. Therefore, the time for calculating the stock price, which consists of the function $\pi(\tau, x)$ at a location $x$ on the right side of equation (12) is unknown. Hence, we had to take the time simply to be the exact number of the trading days on which the stock will be traded as there is no fixed time for the asset price for trading in the frontier markets.

## Conclusion

We have come up with an approach to use Black-Scholes equations to predict the price of any stock of a company of frontier markets including the buying and selling prices of any stock. This approach can be very useful for finding the strike price, volatility and time of expiration of a certain stock as these parameters do not really exist in the frontier markets, which does not trade options. Depending on the stock prices of any trading month, the stock prices of other upcoming months can be predicted and the trend can also easily be known. For doing this, the required parameters of Black-Scholes equation must also need to be changed accordingly, as shown in this paper. Taking the historical stock prices (close price) of a certain month of any year and changing it sequentially with time, the stock prices of later months can be predicted. However, volatility and risk free interest rate will remain constant and known throughout the time of prediction for all the stock prices of a month. It is to be noted that, this process is actually a continuous process. Because, depending on stock prices of a particular month from the past, we get the predicted future stock prices of another month and then depending on the prediction of stock prices of that month, the prediction of stock prices of later months can be achieved. This method can be very useful for traders who trade in frontier markets. The increase and decrease in the volatility can be useful to people for making maximum profit out of the stock as it indicates the trend of the market that is, when the stock price of  a company will increase or decrease. Maximum profit can be made through the idea that, more the volatility, more the price of the stocks. Black-Scholes prediction indicates a price similar to opening price of the day and the original close price is deviated from Black-Scholes prediction because each day continuously trading takes place, as a result of which the stock (close) price also continuously changes. The graphs obtained by data mining to predict the stock prices in Rapidminer are very similar to the graph obtained using Black-Scholes equation. The trend of the market also appears to be very similar for both machine learning approach and Black-Scholes approach, which leads to the conclusion that the Black-Scholes equation can be very useful for calculating and predicting the stock prices of companies of the frontier markets.

**List of the Tables**

| Names Of Companies | Trading Code | Predictive Models and Learners | Attributes Of Dataset |
|---|---|---|---|
| Bangladesh Export Import Company Ltd. | Beximco | Decision Tree | Date |
| Grameenphone Ltd. | GP | Neural Net | Opening Price |
| Olympic Industries Ltd. | Olympic | Ensemble Learning | Close Price |
| ACI Limited | ACI | | High |
| The City Bank Ltd. | City Bank | | Low |
| Adamjee Insurance Co. Ltd. | AICL | | Volume |
| Afric Industries SA | AFI | | |
| ATLANTA | ATL | | |
| BYCO Petroleum Pak Ltd. | BYCO | | |
| C T Holdings PLC | CTHR.N0000 | | |
| Chevron Lubricants Lanka PLC | LLUB.N0000 | | |

Table 1. Names of the companies, predicting methods and models with attributes used for data mining in Rapidminer.

| Predictive Models | Trading Codes Of Companies | Training | Testing |
|---|---|---|---|
| Decision Tree and Neural Net | Beximco | 02.01.1999-31.01.2017 | 01.02.2017-28.02.2017 |
| | ACI | 02.01.1999-30.09.2010 | 03.10.2010-31.10.2010 |
| | GP | 16.11.2009-30.04.2017 | 02.05.2017-31.05.2017 |
| | City Bank | 01.01.2003-31.05.2018 | 03.06.2018-28.06.2018 |
| | Olympic | 22.07.2007-31.05.2018 | 03.06.2018-28.06.2018 |
| | AICL | 04.06.2002-29.03.2013 | 01.04.2013-30.04.2013 |
| | AFI | 05.01.2012-29.02.2016 | 01.03.2016-31.03.2016 |
| | ATL | 22.10.2007-30.03.2018 | 02.04.2018-30.04.2018 |
| | BYCO | 01.08.2002-30.06.2014 | 01.07.2014-28.07.2014 |
| | CTHR.N0000 | 14.03.2000-30.04.2015 | 01.05.2015-29.05.2015 |
| | LLUB.N0000 | 04.01.1999-30.03.2018 | 02.04.2018-30.04.2014 |

Table 2. Training and testing dataset of the companies for decision tree and neural net models.

| Predictive Model | Trading Codes Of Companies | Performance vector |
|---|---|---|
| Decision Tree | Beximco | Root mean squared error: 0.543 +/- 0.000<br>Absolute error: 0.375 +/- 0.392<br>Relative error: 1.13% +/- 1.15%<br>squared error: 0.294 +/- 0.502<br>Correlation: 0.869<br>Squared correlation: 0.756 |
| | ACI | Root mean squared error: 7.033 +/- 0.000<br>Absolute error: 6.021 +/- 3.635<br>Relative error: 1.56% +/- 0.96%<br>Squared error: 49.466 +/- 46.068<br>Correlation: 0.478<br>Squared correlation: 0.228 |
| | GP | Root mean squared error: 1.884 +/- 0.000<br>Absolute error: 1.358 +/- 1.306<br>Relative error: 0.41% +/- 0.39%<br>Squared error: 3.549 +/- 5.877<br>Correlation: 0.921<br>Squared correlation: 0.849 |
| | City Bank | Root mean squared error: 0.601 +/- 0.000<br>Absolute error: 0.478 +/- 0.364<br>Relative error: 1.45% +/- 1.09%<br>Squared error: 0.362 +/- 0.616<br>Correlation: 0.880<br>Squared correlation: 0.774 |
| | Olympic | Root mean squared error: 1.978 +/- 0.000<br>Absolute error: 1.527 +/- 1.258<br>Relative error: 0.66% +/- 0.55%<br>Squared error: 3.913 +/- 5.941<br>Correlation: 0.945<br>Squared correlation: 0.894 |
| | AICL | Root mean squared error: 1.316 +/- 0.000<br>Absolute error: 0.905 +/- 0.955<br>Relative error: 1.20% +/- 1.30%<br>Squared error: 1.732 +/- 3.572<br>Correlation: 0.922<br>Squared correlation: 0.849 |
| | AFI | Root mean squared error: 4.877 +/- 0.000<br>Absolute error: 2.170 +/- 4.368<br>Relative error: 0.64% +/- 1.28%<br>Squared error: 23.788 +/- 64.144<br>Correlation: 0.862<br>Squared correlation: 0.743 |
| | ATL | Root mean squared error: 0.826 +/- 0.000<br>Absolute error: 0.677 +/- 0.472<br>Relative error: 0.86% +/- 0.60%<br>Squared error: 0.682 +/- 0.760<br>Correlation: 0.726<br>Squared correlation: 0.527 |
| | BYCO | Root mean squared error: 0.096 +/- 0.000 |

| | | Absolute error: 0.068 +/- 0.067 |
| | | Relative error: 0.61% +/- 0.61% |
| | | Squared error: 0.009 +/- 0.018 |
| | | Correlation: 0.886 |
| | | Squared correlation: 0.785 |
| | CTHR.N0000 | Root mean squared error: 1.352 +/- 0.000 |
| | | Absolute error: 0.489 +/- 1.260 |
| | | Relative error: 0.42% +/- 1.08% |
| | | Squared error: 1.827 +/- 7.398 |
| | | Correlation: 0.647 |
| | | Squared correlation: 0.419 |
| | LLUB.N0000 | Root mean squared error: 4.033 +/- 0.000 |
| | | Absolute error: 3.231 +/- 2.414 |
| | | Relative error: 3.29% +/- 2.45% |
| | | Squared error: 16.264 +/- 16.877 |
| | | Correlation: 0.850 |
| | | Squared correlation: 0.722 |

Table 3. Performance vector values obtained from decision tree model using Rapidminer.

| Predictive Model | Trading Codes Of Companies | Performance vector |
| --- | --- | --- |
| Neural Net | Beximco | Prediction trend accuracy: 0.457 +/- 0.363 (micro average: 0.457) |
| | ACI | Prediction trend accuracy: 0.452 +/- 0.354 (micro average: 0.452) |
| | GP | Prediction trend accuracy: 0.468 +/- 0.357 (micro average: 0.468) |
| | City Bank | Prediction trend accuracy: 0.460 +/- 0.358 (micro average: 0.460) |
| | Olympic | Prediction trend accuracy: 0.481 +/- 0.360 (micro average: 0.481) |
| | AICL | Prediction trend accuracy: 0.453 +/- 0.352 (micro average: 0.453) |
| | AFI | Prediction trend accuracy: 0.440 +/- 0.370 (micro average: 0.440 |
| | ATL | Prediction trend accuracy: 0.513 +/- 0.348 (micro average: 0.513) |
| | BYCO | Prediction trend accuracy: 0.473 +/- 0.359 (micro average: 0.473) |
| | CTHR.N0000 | Prediction trend accuracy: 0.446 +/- 0.355 (micro average: 0.446) |
| | LLUB.N0000 | Prediction trend accuracy: 0.401 +/- 0.356 (micro average: 0.401) |

Table 4. Performance vector values obtained from neural net model using Rapidminer.

| Predictive method | Trading Codes Of Companies | Training |
|---|---|---|
| Ensemble learner | Beximco | 02.01.1999-28.02.2017 |
| | ACI | 02.01.1999-31.10.2010 |
| | GP | 16.11.2009-31.05.2017 |
| | City Bank | 01.01.2003-28.06.2018 |
| | Olympic | 22.07.2007-28.06.2018 |
| | AICL | 04.06.2002-30.04.2013 |
| | AFI | 05.01.2012-31.03.2016 |
| | ATL | 22.10.2007-30.04.2018 |
| | BYCO | 01.08.2002-28.07.2014 |
| | CTHR.N0000 | 14.03.2000-29.05.2015 |
| | LLUB.N0000 | 04.01.1999-30.04.2018 |

Table 5. Training dataset of the companies for ensemble learning method.

| Predictive Method | Trading Codes Of Companies | Performance vector |
|---|---|---|
| Ensemble learning method | Beximco | Root mean squared error: 0.452 +/- 0.000<br>Absolute error: 0.228 +/- 0.391<br>Relative error: 0.31% +/- 0.43%<br>Squared error: 0.205 +/- 0.756<br>Correlation: 1.000<br>Squared correlation: 1.000 |
| | ACI | Root mean squared error: 8.550 +/- 0.000<br>Absolute error: 6.192 +/- 5.896<br>Relative error: 1.45% +/- 1.20%<br>Squared error: 73.099 +/- 81.082<br>Correlation: 1.000<br>Squared correlation: 1.000 |
| | GP | Root mean squared error: 0.482 +/- 0.000<br>Absolute error: 0.105 +/- 0.471<br>Relative error: 0.04% +/- 0.16%<br>Squared error: 0.233 +/- 3.557<br>Correlation: 1.000<br>Squared correlation: 1.000 |
| | City Bank | Root mean square error: 0.321 +/- 0.000<br>Absolute error: 0.205 +/- 0.247<br>Relative error: 0.94% +/- 0.82%<br>Squared error: 0.103 +/- 0.789<br>Correlation: 1.000<br>Squared correlation: 0.999 |
| | Olympic | Root mean squared error: 0.473 +/- 0.000<br>Absolute error: 0.268 +/- 0.389 |

| | | Relative error: 0.10% +/- 0.13%<br>Squared error: 0.224 +/- 3.330<br>Correlation: 1.000<br>Squared correlation: 1.000 |
|---|---|---|
| | AICL | Root mean squared error: 0.044 +/- 0.000<br>Absolute error: 0.032 +/- 0.030<br>Relative error: 0.04% +/- 0.04%<br>Squared error: 0.002 +/- 0.004<br>Correlation: 1.000<br>Squared correlation: 1.000 |
| | AFI | Root mean squared error: 0.789 +/- 0.000<br>Absolute error: 0.578 +/- 0.537<br>Relative error: 0.17% +/- 0.16%<br>Squared error: 0.622 +/- 0.740<br>Correlation: 1.000<br>Squared correlation: 1.000 |
| | ATL | Root mean squared error: 0.063 +/- 0.000<br>Absolute error: 0.047 +/- 0.042<br>Relative error: 0.09% +/- 0.09%<br>Squared error: 0.004 +/- 0.006<br>Correlation: 1.000<br>Squared correlation: 1.000 |
| | BYCO | Root mean squared error: 0.006 +/- 0.000<br>Absolute error: 0.004 +/- 0.004<br>Relative error: 0.04% +/- 0.04%<br>Squared error: 0.000 +/- 0.000<br>Correlation: 1.000<br>Squared correlation: 1.000 |
| | CTHR.N0000 | Root mean squared error: 0.149 +/- 0.000<br>Absolute error: 0.122 +/- 0.086<br>Relative error: 0.09% +/- 0.07%<br>Squared error: 0.022 +/- 0.036<br>Correlation: 1.000<br>Squared correlation: 1.000 |
| | LLUB.N0000 | Root mean squared error: 1.523 +/- 0.000<br>Absolute error: 1.170 +/- 0.975<br>Relative error: 0.72% +/- 0.51%<br>Squared error: 2.319 +/- 2.748<br>Correlation: 1.000<br>Squared correlation: 1.000 |

Table 6. Performance vector values obtained from ensemble learning method

| Date | Close Price | Daily Price Change(X) | Mean |
|---|---|---|---|
| 29-02-16 | 339.5 | 0.002949855 | 0.001640969 |

| 26-02-16 | 338.5 | 0.002958582 | |
|----------|-------|-------------|--|
| 25-02-16 | 337.5 | 0.00445435 | |
| 24-02-16 | 336 | -0.00445435 | |
| 23-02-16 | 337.5 | 0.002967361 | |
| 22-02-16 | 336.5 | 0.002976193 | |
| 19-02-16 | 335.5 | 0.005979091 | |
| 18-02-16 | 333.5 | 0.003003005 | |
| 17-02-16 | 332.5 | 0.00301205 | |
| 16-02-16 | 331.5 | -0.00900907 | |
| 15-02-16 | 334.5 | 0.00449439 | |
| 12-02-16 | 333 | 0.009049836 | |
| 11-02-16 | 330 | 0 | |
| 10-02-16 | 330 | 0.004555817 | |
| 09-02-16 | 328.5 | 0.006106889 | |
| 08-02-16 | 326.5 | -0.009146405 | |
| 05-02-16 | 329.5 | 0 | |
| 04-02-16 | 329.5 | -0.001516301 | |
| 03-02-16 | 330 | 0 | |
| 02-02-16 | 330 | 0.001516301 | |
| 01-02-16 | 329.5 | 0.004562746 | |
| 29-01-16 | 328 | | |

Table 7. Calculation of daily price change and mean of company AFI of the month February 2016 used for calculating stock prices of the month March 2016.

| Black-Scholes-Merton Equation | |
|-------------------------------|--|
| Input | Values |
| Current stock price (S) [close price of 1st February, 2016] | 329.5 |
| Mean or Daily Price Change [see table 7] | 0.001640969 |
| e^x [see equation (6)] | 1.001642316 |
| k/S [see equation (6)] | 1.001642316 |
| Exercise price/Strike Price (k) [see equation (7)] | 330.0411432 |
| Time to maturity of option (T) [see equation (8)] | 0.164 |
| Risk-free rate of interest (r) [see table 11] | 0.0248 |
| Dividend yield (q) [see table 10] | 5.89% |
| Stock volatility (Sigma) [see table 9] | 56.864 |
| d1 [see equation (3)] | 11.51376833 |
| d2 [see equation (4)] | -11.51439655 |
| N(d1) | 1.0000 |
| N(d2) | 0.0000 |
| N(-d1) | 0.0000 |
| N(-d2) | 1.0000 |
| Price of call option (C) (Right To Buy) [see equation 1] | 326.3324849 |

| Price of put option (P) (Right To Sell) [equation 2] | 328.7015259 |
| Average of call option and put option (close Price of 16 March, 2016) | 327.5170054 |

Table 8. Calculation of call option and put option to find the buying and selling price of the stock of the company Afric Industries SA of the 1$^{st}$ March, 2016 depending on close price of 1$^{st}$ February, 2016

| Month | Volatility (Standard Deviation, sigma) |
| March 2015 | 6.464 |
| April 2015 | 7.429 |
| May 2015 | 8.472 |
| June 2015 | 5.543 |
| July 2015 | 1.586 |
| August 2015 | 2.502 |
| September 2015 | 1.451 |
| October 2015 | 3.499 |
| November 2015 | 4.566 |
| December 2015 | 3.738 |
| January 2016 | 7.871 |
| February 2016 | 3.743 |
| Sum | 56.864 |

Table 9. Volatility of one year obtained from MATLAB for calculating stock price of company AFI of the month of March 2016.

| Company Names | Stock Exchange | Source of dividend yield | Date | Dividend Yield |
| --- | --- | --- | --- | --- |
| Beximco | Dhaka Stock Exchange | Thomson Reuter | 31.01.2017 | 0% |
| ACI | Dhaka Stock Exchange | Thomson Reuter | 30.06.2010 | 1.3% |
| GP | Dhaka Stock Exchange | Thomson Reuter | 28.04.2017 | 5.18% |
| City Bank | Dhaka Stock Exchange | Thomson Reuter | 31.05.2018 | 2.14% |
| Olympic | Dhaka Stock Exchange | Thomson Reuter | 31.05.2018 | 1.9% |
| AICL | Pakistan Stock Exchange | Thomson Reuter | 29.03.2013 | 3.56% |
| AFI | Casablanca Stock Exchange | Thomson Reuter | 29.02.2016 | 5.89% |
| ATL | Casablanca Stock Exchange | Thomson Reuter | 30.03.18 | 3.5% |
| BYCO | Pakistan Stock Exchange | Thomson Reuter | | |

| CTHR.N0000 | Colombo Stock Exchange | Thomson Reuter | 30.04.2015 | 2.57% |
| LLUB.N0000 | Colombo Stock Exchange | Thomson Reuter | 30.03.2018 | 11.72% |

Table 10. Stocks with dividend yield and the respective companies are shown in the figure with date, which are obtained using Thomson Reuter.

| Company Names | Stock Exchange | Source of treasury bill rate | Date | Treasury Bill Rate |
|---|---|---|---|---|
| Beximco | Dhaka Stock Exchange | Bangladesh Bank (Thomson Reuter) | 31.01.2017 | 3 month treasury bill rate (2.97%) |
| ACI | Dhaka Stock Exchange | Bangladesh Bank (Thomson Reuter) | 30.06.2010 | 3 month treasury bill rate (2.42%) |
| GP | Dhaka Stock Exchange | Bangladesh Bank (Thomson Reuter) | 28.04.2017 | 3 month treasury bill rate (2.86%) |
| City Bank | Dhaka Stock Exchange | Bangladesh Bank (Thomson Reuter) | 31.05.2018 | 3 month treasury bill rate (0.86%) |
| Olympic | Dhaka Stock Exchange | Bangladesh Bank (Thomson Reuter) | 31.05.2018 | 3 month treasury bill rate (0.86%) |
| AICL | Pakistan Stock Exchange | State Bank Of Pakistan (Thomson Reuter) | 15.03.2013 | 6 month treasury bill rate (9.42%) |
| AFI | Casablanca Stock Exchange | Bank Al-Maghrib (Thomson Reuter) | 29.02.2016 | 13 month treasury bill rate (2.48%) |
| ATL | Casablanca Stock Exchange | Bank Al-Maghrib (Thomson Reuter) | 30.03.2018 | 13 month treasury bill rate (2.16%) |
| BYCO | Pakistan Stock Exchange | State Bank Of Pakistan (Thomson Reuter) | 15.06.2014 | 6 month treasury bill rate (9.97%) |
| CTHR.N0000 | Colombo Stock Exchange | Central Bank Of Sri Lanka (Thomson Reuter) | 15.04.2015 | 3 month treasury bill rate (6.1%) |
| LLUB.N0000 | Colombo Stock Exchange | Central Bank Of Sri Lanka (Thomson Reuter) | 15.03.2018 | 3 month treasury bill rate (8.15%) |

Table 11. Treasury bill rates are shown, which are taken as the risk free interest rates for the Black-Scholes equation. The company names and corresponding stock exchanges with date are shown in the figure and the data is obtained using software Thomson Reuter.

# Figures and Captions List

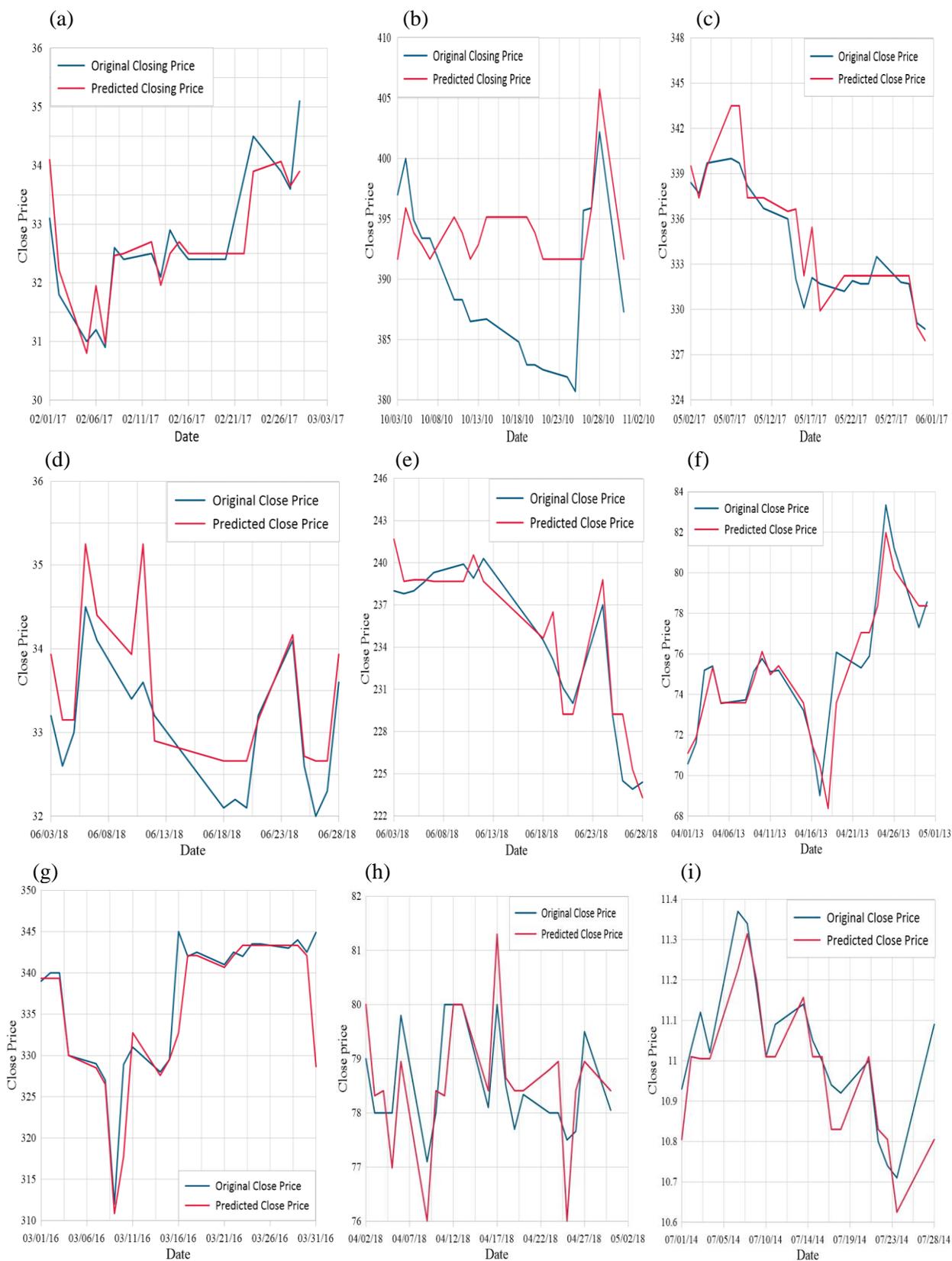

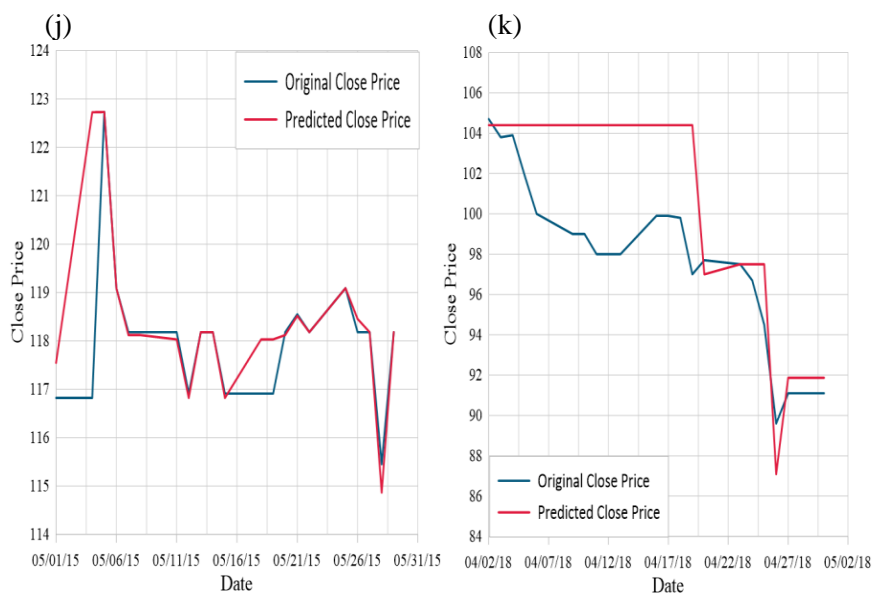

Figure 2. Comparison between original and predicted close price obtained from Rapidminer using decision tree model (a) for the month February 2017 of company Beximco (b) for the month October 2010 of company ACI. (c) for the month May 2017 of company GP. (d) for the month June 2018 of company City Bank. (e) for the month June 2018 of company Olympic. (f) for the month April 2013 of company AICL. (g) for the month March 2016 of company AFI. (h) for the month April 2018 of company ATL. (i) for the month July 2014 of company BYCO. (j) for the month April 2015 of company CTHR.N0000. (k) for the month April 2015 of company LLUB.N0000.

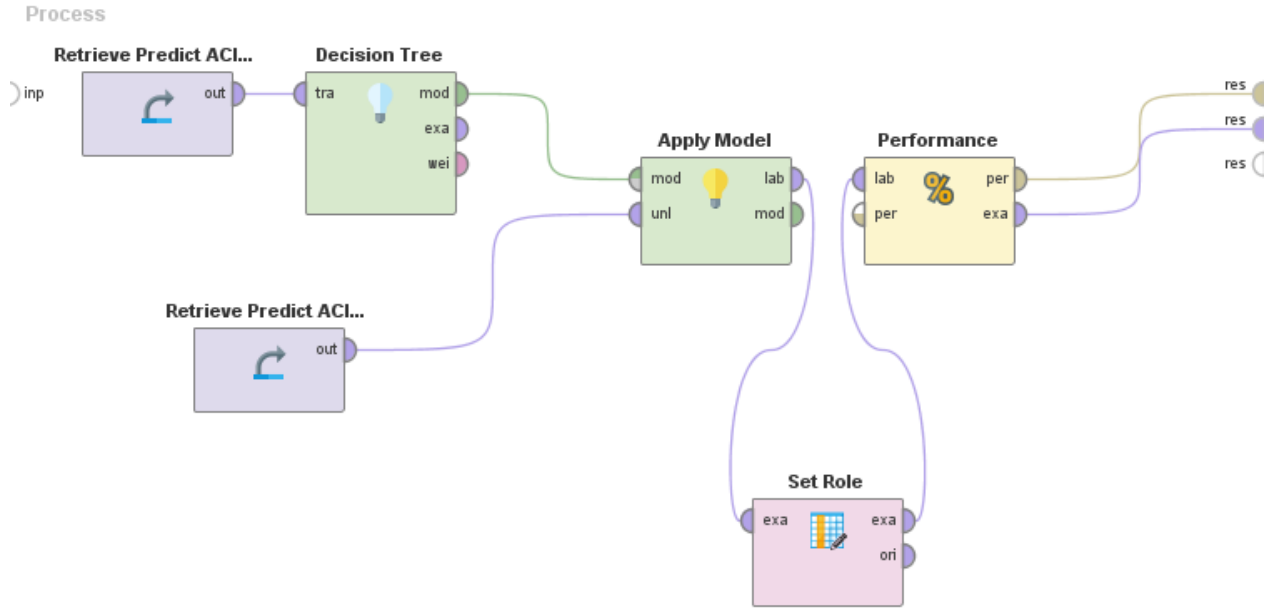

Figure 3. Decision tree model for predicting stock prices of the companies.

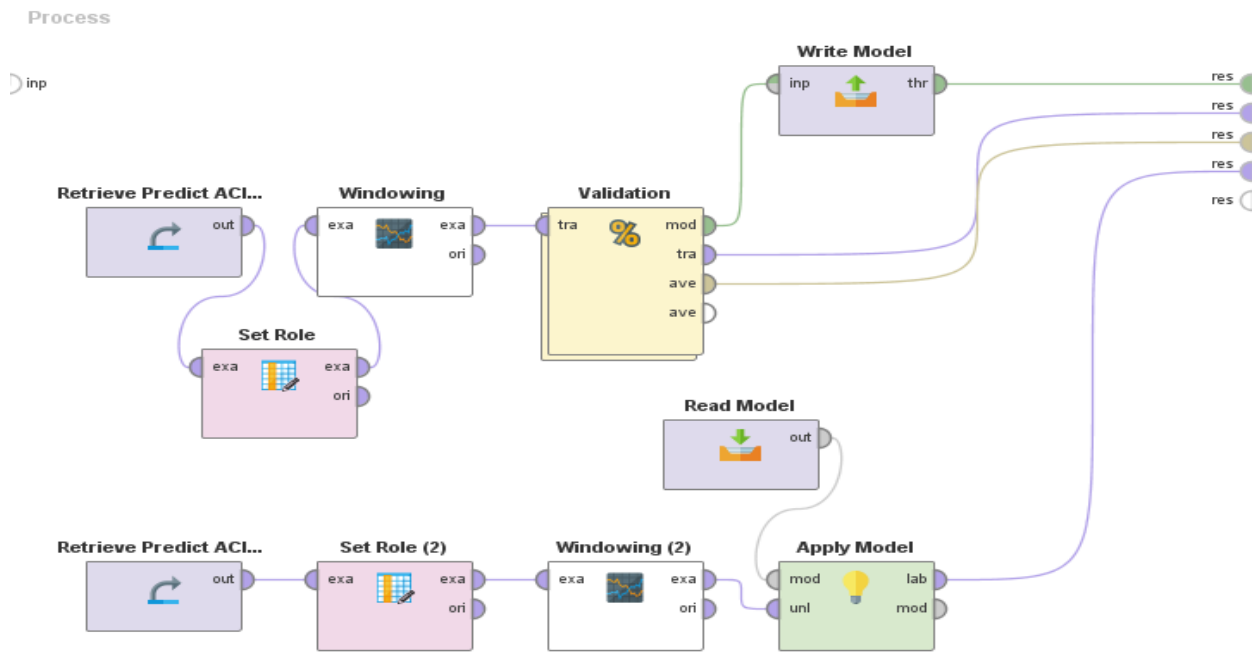

Figure 4. Neural net model for predicting stock prices of the companies.

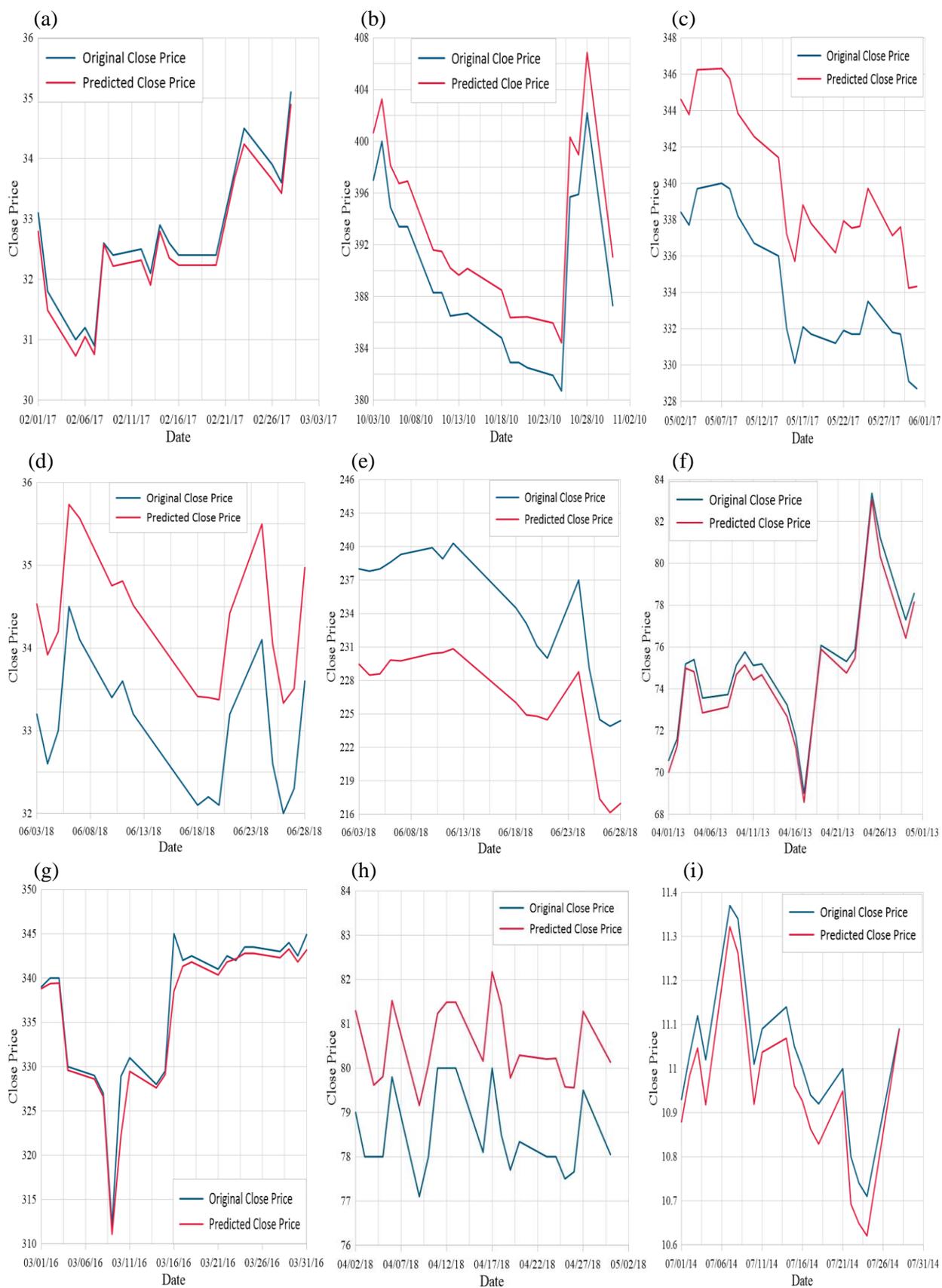

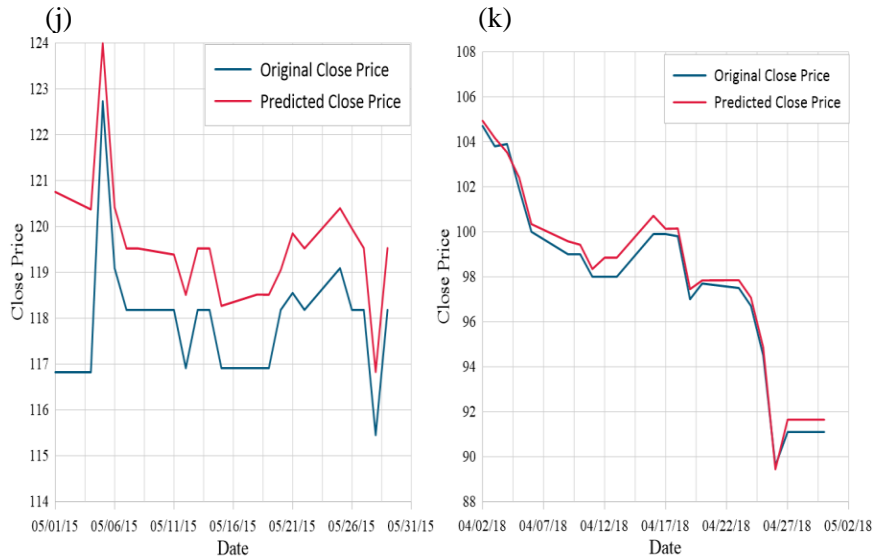

Figure 5. Comparison between original and predicted close price obtained from Rapidminer using neural net model (a) for the month February 2017 of company Beximco. (b) for the month October 2010 of company ACI. (c) for the month May 2017 of company GP. (d) for the month June 2018 of company City Bank. (e) for the month June 2018 of company Olympic. (f) for the month April 2013 of company AICL. (g) for the month March 2016 of company AFI. (h) for the month April 2018 of company ATL. (i) for the month July 2014 of company BYCO. (j) for the month May 2015 of company CTHR.N0000. (k) for the month April 2015 of company LLUB.N0000.

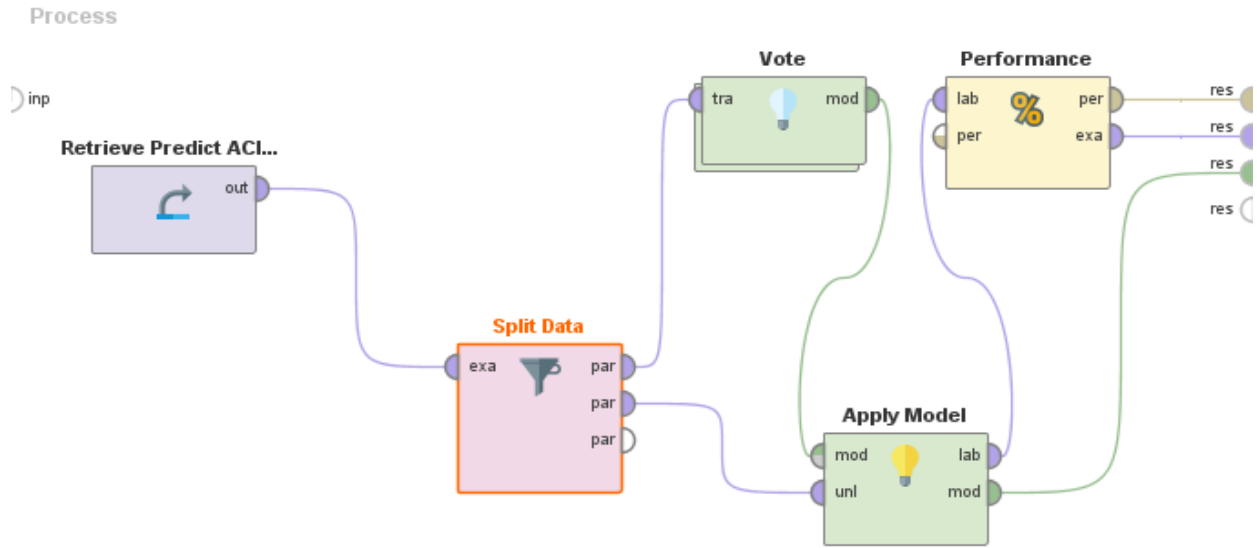

Figure 6. Ensemble learning model for predicting the close price in Rapidminer.

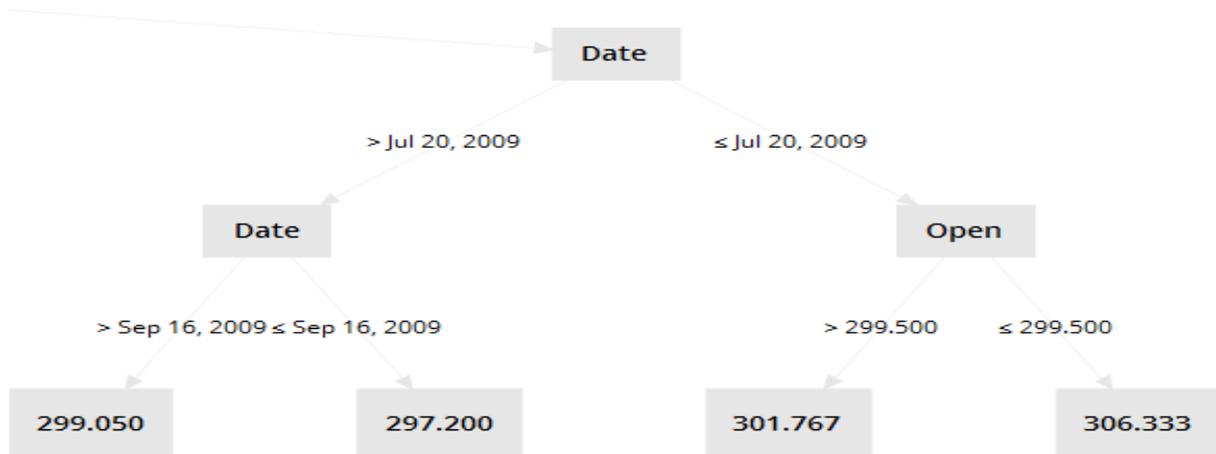

Figure 7. Decision tree leaf used in ensemble learning method.

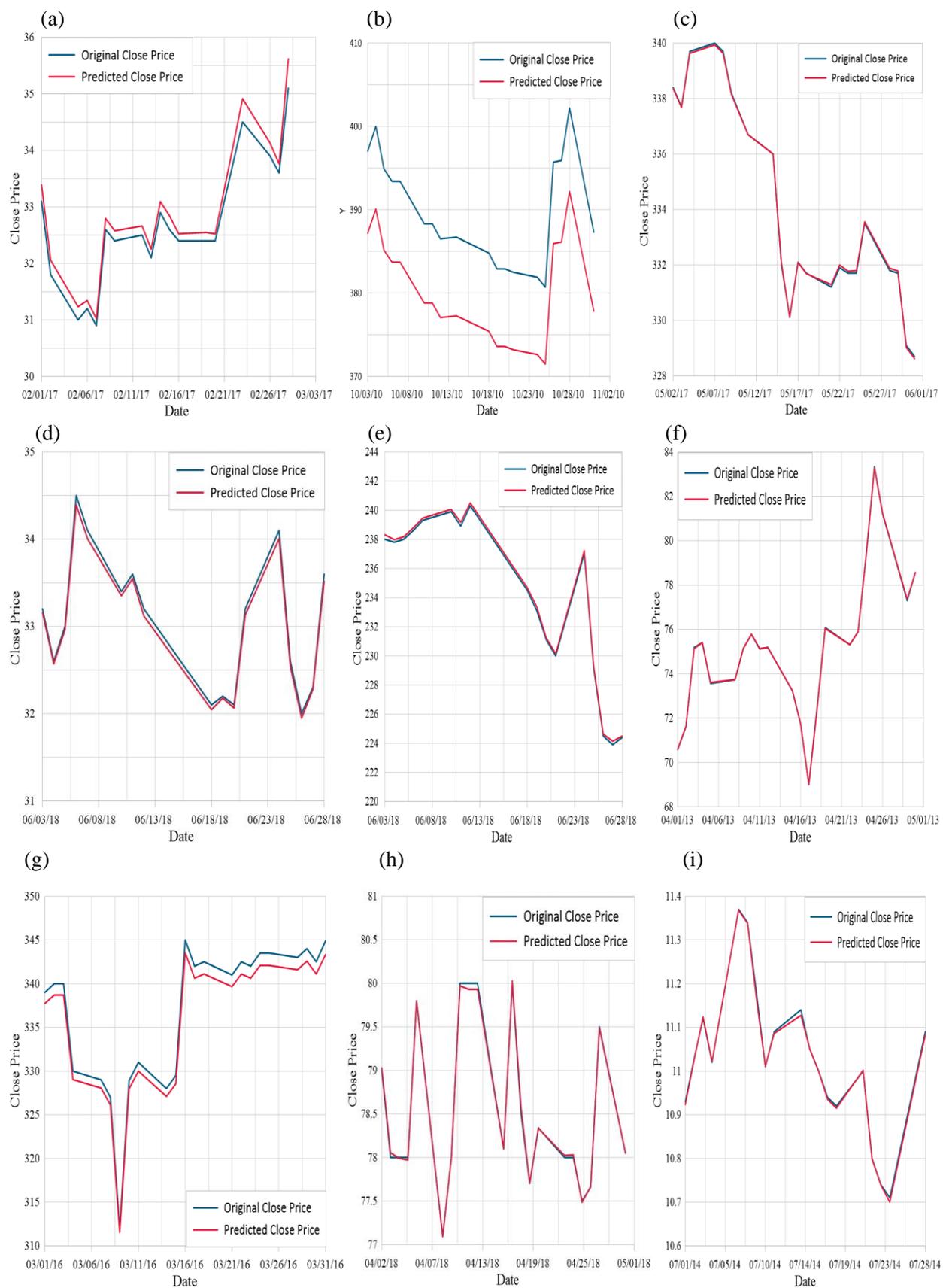

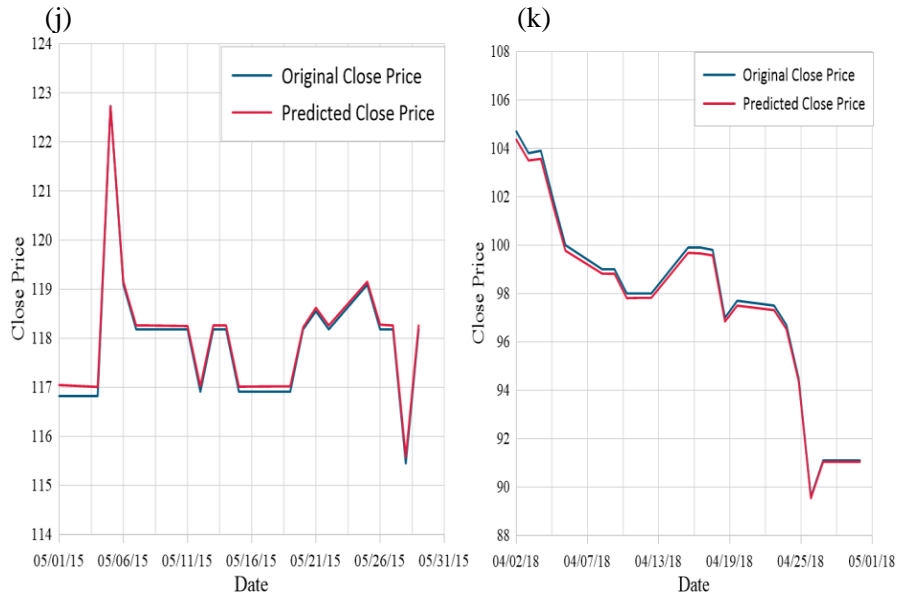

Figure 8. Comparison between original and predicted close price obtained from Rapidminer using ensemble learning method (a) for the month February 2017 of company Beximco. (b) for the month October 2010 of company ACI. (c) for the month May 2017 of company GP. (d) for the month June 2018 of company City Bank. (e) for the month June 2018 of company Olympic. (f) for the month April 2013 of company AICL. (g) for the month March 2016 of company AFI. (h) for the month April 2018 of company ATL. (i) for the month July 2014 of company BYCO. (j) for the month April 2015 of company CTHR.N0000. (k) for the month April 2015 of company LLUB.N0000.

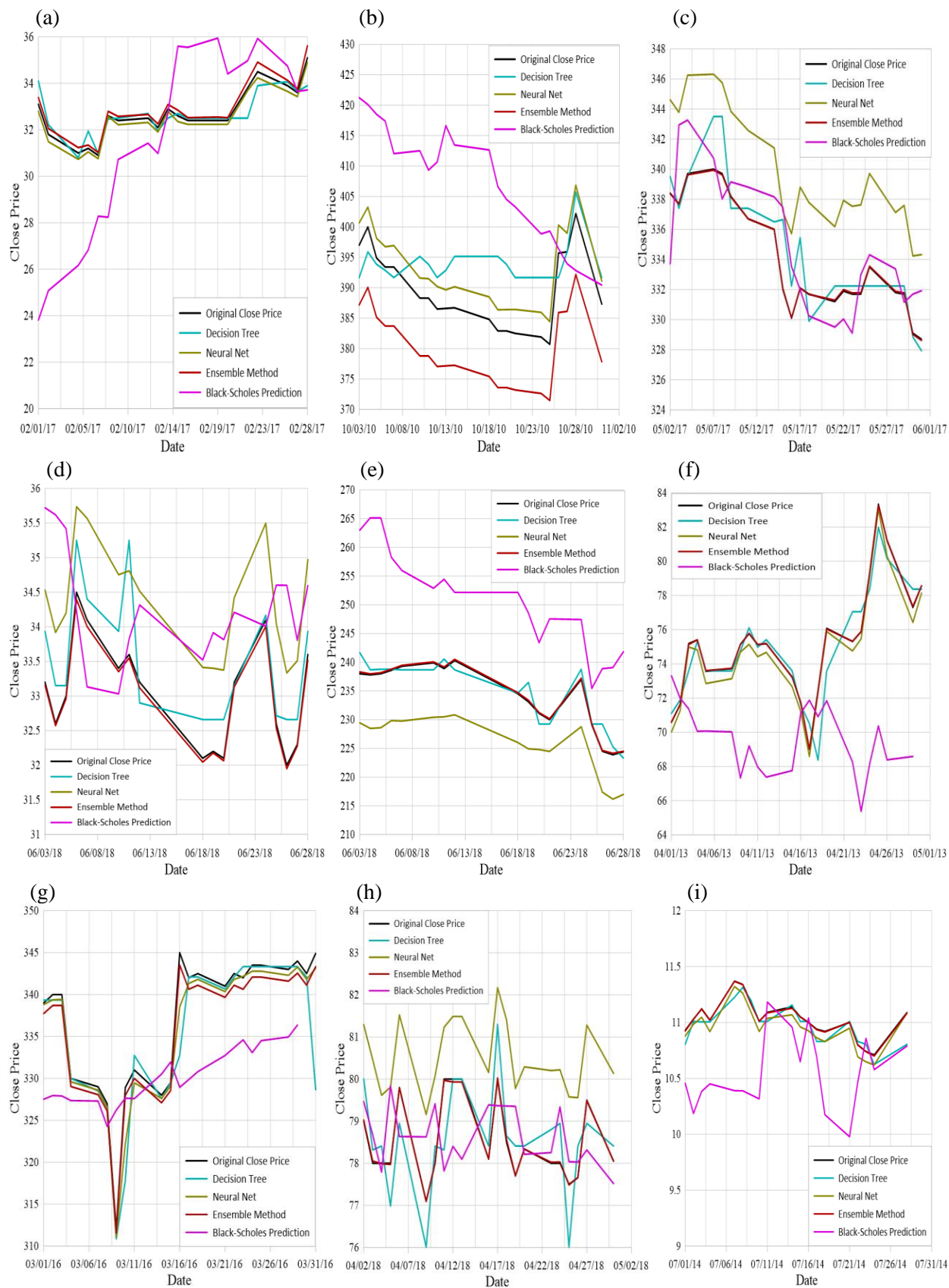

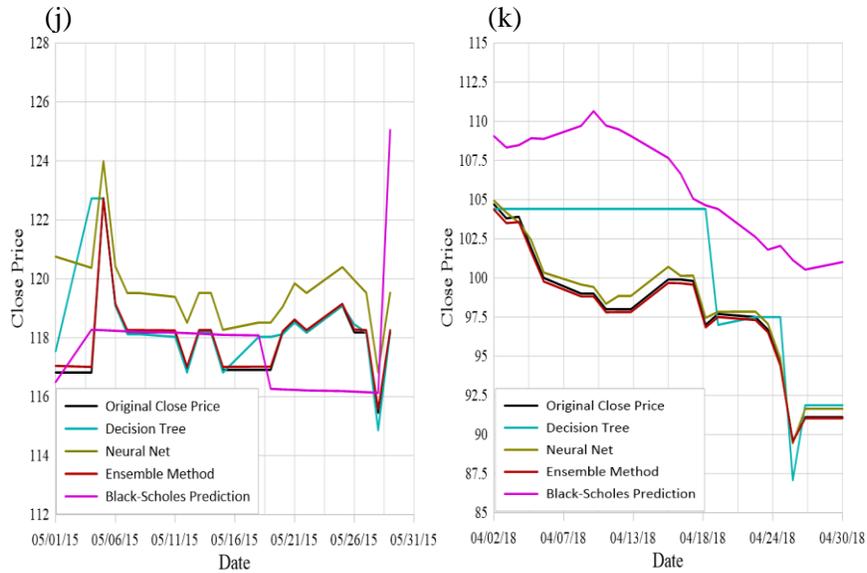

Figure 9. Comparison between original close price and predicted close price by BSOPM and Rapidminer (a) of company Beximco of the month February 2017. (b) of company ACI of the month October 2010. (c) of company GP of the month May 2107. (d) of company City Bank of the month June 2018. (e) of company Olympic of the month June 2018. (f) of company AICL of the month April 2013. (g) of company AFI of the month March 2016. (h) of company ATL of the month April 2018. (i) of company BYCO of the month July 2014. (j) of company CTHR.N0000 of the month May 2015. (k) of company LLUB.N0000 of the month April 2018.

# Supplement Information of "Predicting the Stock price of Frontier Markets Using modified Black-Scholes Option Pricing Model and Machine Learning"

Reaz Chowdhury[1], M.R.C. Mahdy[1,2*], Tanisha Nourin Alam[1], Golam Dastegir Al Quaderi[3]

[1]*Department of Electrical & Computer Engineering, North South University, Bashundhara, Dhaka 1229, Bangladesh*

[2]*Pi Labs Bangladesh LTD, ARA Bhaban, 39, Kazi Nazrul Islam Avenue, Kawran Bazar, Dhaka 1215, Bangladesh*

[3]*Department of Physics, University of Dhaka, Dhaka 1000, Bangladesh*

*Corresponding Author: mahdy.chowdhury@northsouth.edu

**S1: MATLAB code to find the volatility of a stock**

```
[CLOSEP,DATE,RAWDATA]=xlsread('2.xlsx','Sheet1');

CLOSEP2 = CLOSEP*1;

dates=DATE(2:end,1);

dates_num = datenum(dates,'dd-mm-yyyy');

dates_vec = datevec(dates_num);

X = dates_vec(:,3);

Y= CLOSEP2;

plot(X,Y);

title('PLOT OF CLOSEP vs DAYS')

xlabel('DAYS')

ylabel('CLOSEP')

grid
```

Figure 1s MATLAB code to find the volatility of a company's stock of one month and then volatilities of all twelve months are summed together to give as input into the Black-Scholes equation.

**S2: A sample of neural network**

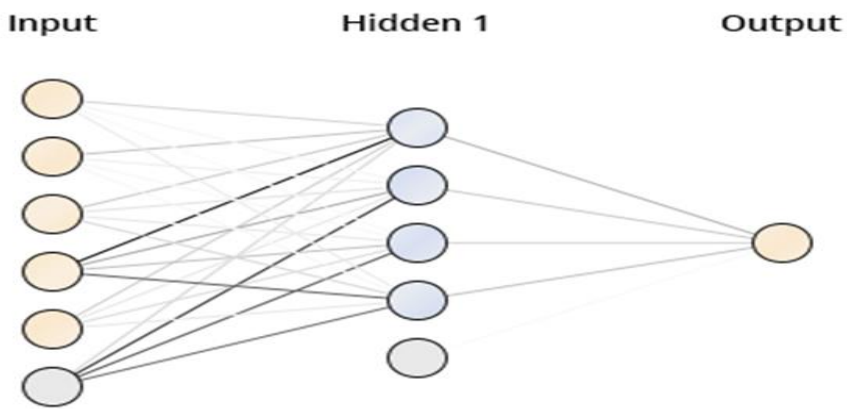

Figure 2s A sample of improved neural network from Rapidminer.

**S3: Comparison between original and predicted close prices of all eleven companies using Black-Scholes equations**

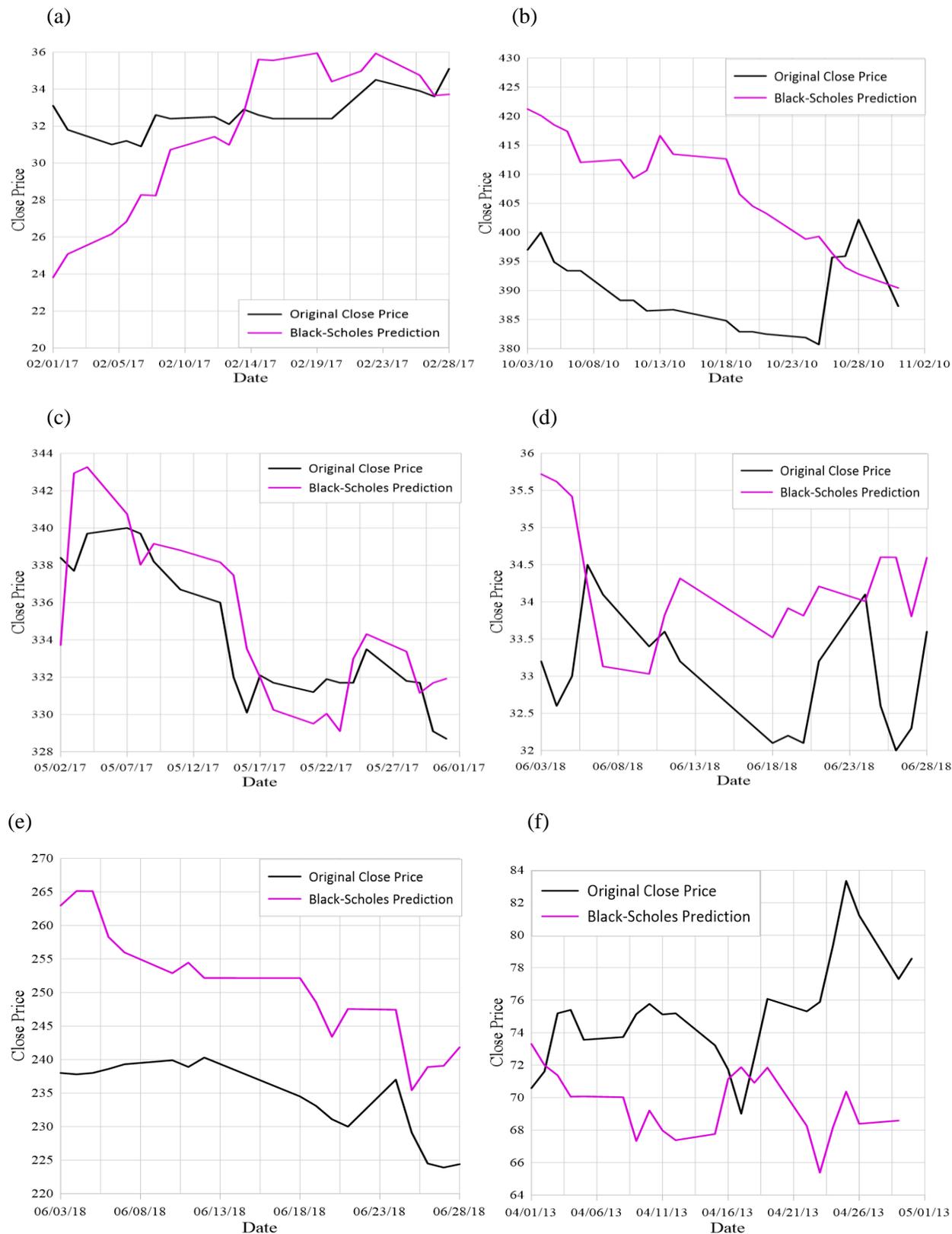

(g)

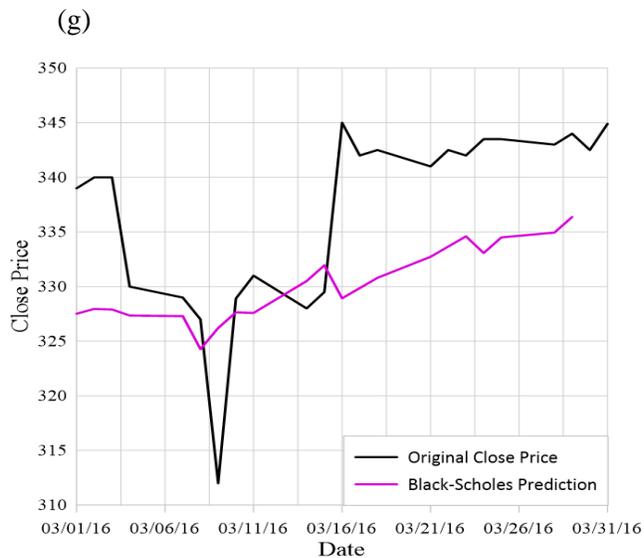

(h)

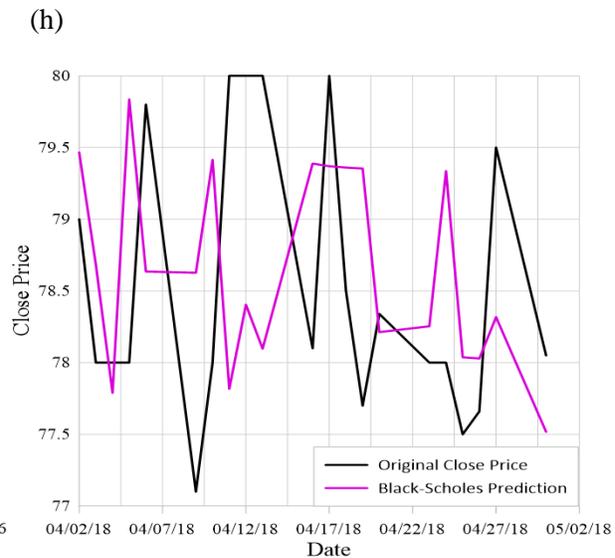

(i)

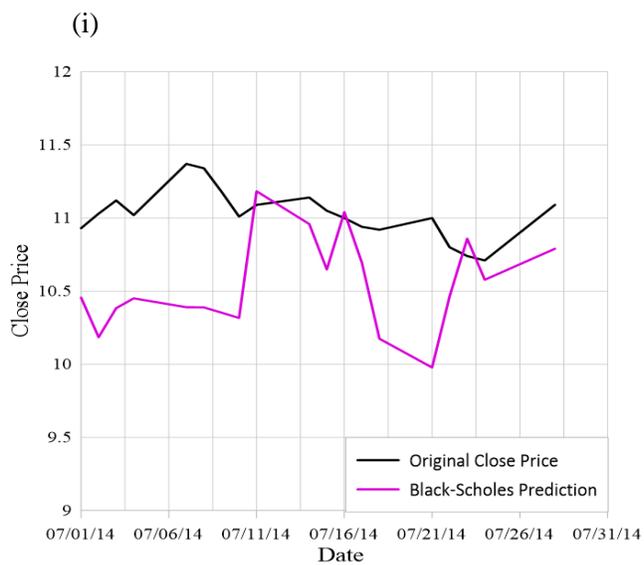

(j)

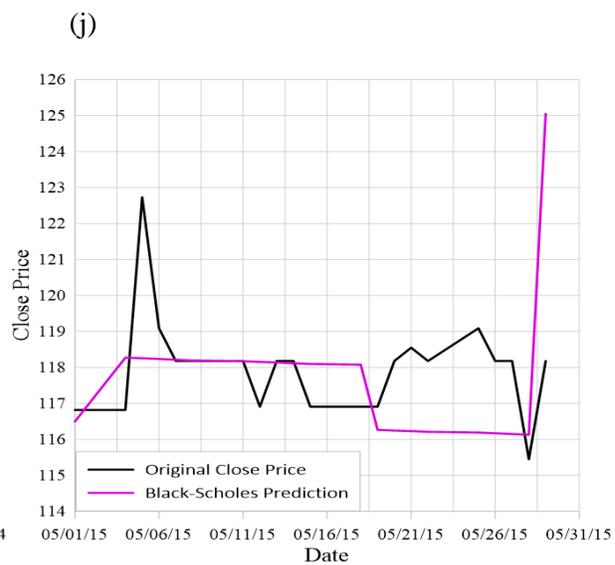

(k)

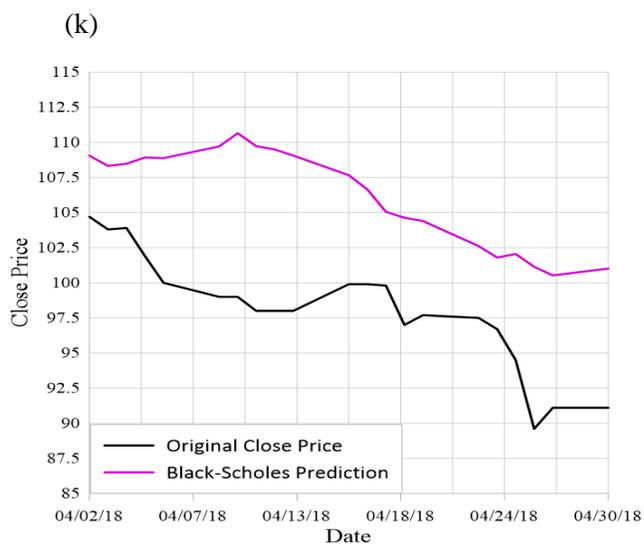

Figure 3s Comparison between original close price and predicted close price using Black-Scholes equations (a) of company Beximco of the month February 2017 depending on January 2017. (b) of company ACI of the month October 2010 depending on June 2010. (c) of company GP of the month May 2107 depending on April 2017. (d) of the month June 2018 depending on May 2108. (e) of company Olympic of the month June 2018 depending on May 2018. (f) of the month April 2013 depending on March 2103. (g) of company AFI of the month March 2016 depending on February 2016. (h) of company ATL of the month April 2018 depending on March 2018. (i) of company BYCO of the month July 2014 depending on June 2014. (j) of company CTHR.N0000 of the month May 2015 depending on April 2015. (k) of company LLUB.N0000 of the month April 2018 depending on March 2018.

**S4: Reference of data taken from various websites for Dhaka Stock Exchange, Pakistan Stock Exchange, Casablanca Stock Exchange and Colombo Stock Exchange**

| Company Names and Trading Code | Stock Exchange | Source of Data |
|---|---|---|
| Bangladesh Export Import Company Ltd. (Beximco) | Dhaka Stock Exchange | Dhaka Stock Exchange |
| ACI Limited (ACI) | Dhaka Stock Exchange | Dhaka Stock Exchange |
| Grameenphone Ltd. (GP) | Dhaka Stock Exchange | Dhaka Stock Exchange |
| The City Bank Ltd. (City Bank) | Dhaka Stock Exchange | Dhaka Stock Exchange |
| Olympic Industries Ltd. (Olympic) | Dhaka Stock Exchange | Dhaka Stock Exchange |
| Adamjee Insurance Co. Ltd. (AICL) | Pakistan Stock Exchange | StataProfessor |
| Afric Industries SA (AFI) | Casablanca Stock Exchange | Thomson Reuter |
| ATLANTA (ATL) | Casablanca Stock Exchange | Thomson Reuter |
| BYCO Petroleum Pak Ltd. (BYCO) | Pakistan Stock Exchange | StataProfessor |
| City Holdings PLC (CTHR.N0000) | Colombo Stock Exchange | Thomson Reuter |
| Chevron Lanka PLC (LLUB.N0000) | Colombo Stock Exchange | Thomson Reuter |

Figure 4s Sources of all collected data depending on which this research has been done and after collection of data from these various sites for any missing values, which falls under any trading date, the entire row is removed using Microsoft excel and then the rest are taken in dataset.